\documentclass[10pt,conference]{IEEEtran}
\usepackage{blindtext, graphicx}
\usepackage{color}
\usepackage{subcaption}
\usepackage{listings}
\usepackage{amssymb}
\usepackage{multirow}
\usepackage{etoolbox}
\usepackage{algorithm}
\usepackage{syntax}
\usepackage{comment}
\usepackage{hyperref}
\usepackage{amsthm}
\usepackage{amsmath}
\usepackage{xspace}
\usepackage[svgnames]{xcolor}
\usepackage{tcolorbox}
\usepackage[all]{nowidow}
\usepackage{booktabs}
\usepackage{algorithmic}

\theoremstyle{definition}

\newtheorem{example}{Example}

\let\labelindent\relax
\usepackage{enumitem}
\setlist{itemsep=3pt}

\usepackage{pifont}
\newcommand{\cmark}{\ding{51}}%
\newcommand{\xmark}{\ding{55}}%
\newcommand{\starmark}{\ding{72}}%

\newtoggle{release}
\newcommand{\draftonly}[1]{\iftoggle{release}{}{#1}}
\togglefalse{release}
\toggletrue{release}

\newcommand {\loris}[1]{\draftonly{{\color{blue}\bf{L: #1}\normalfont}}}

\newcommand {\gustavo}[1]{\draftonly{{\color{orange}\bf{G: #1}\normalfont}}}

\newcommand{\paraheading}[1]{\noindent \textbf{\textit{#1}} \hspace*{0em}}
\newcommand*{\kind}[1]{\tcbox[size=tight, boxsep=0.3mm, box align=base, nobeforeafter, fontupper=\small]{\text{\code{\vphantom{|}#1}}}}

\newcommand{\numberOfRepetitiveChanges}{59}
\newcommand{\numberOfChangesRoslyn}{27}
\newcommand{\numberOfChangesEntityFramework}{15}
\newcommand{\numberOfChangesNuGet}{17}
\newcommand{\medianPerChange}{5}
\newcommand{\numberOfMultipleFiles}{15}
\newcommand{\percentOfMultipleFiles}{25}
\newcommand{\synthesizedCases}{58}
\newcommand{\numberOfNeutralChanges}{11}
\newcommand{\numberOfConfirmed}{1}

\newcommand{\numberOfIncorrect}{10}
\newcommand{\numberOfNonIdentical}{40}
\newcommand{\percentOfNonIdentical}{68}
\newcommand{\numberOfRepetitiveChangesEqualsCommit}{38}
\newcommand{\percentOfRepetitiveChangesEqualsCommit}{64}
\newcommand{\numberOfRepetitiveChangesMoreCommit}{21}

\newcommand{\totalNumberOfCommits}{404}

\newcommand{\examplesToEdit}{2.8}
\newcommand{\maxNumberOfLocations}{60}

\newcommand{\code}[1]{\foops{#1}}
\newcommand{\foops}[1]{\mbox{\texttt{#1}}}
\newcommand{\refactoringAccuracy}{83}
\newcommand{\prose}{PROSE\xspace}
\newcommand{\totalattempts}{21,781}
\newcommand{\totalstudents}{720}
\newcommand{\totaltasks}{4}
\newcommand{\averageHelpedStudents}{87}
\newcommand{\averageHelpedStudentsIncremental}{44}
\newcommand{\averageAttemptsToFix}{5.2}
\newcommand{\averageAttemptsToComplete}{8.7}

\newcommand{\averageAttemptsToFixIncremental}{6.8}

\newcommand{\technique}{\textsc{Refazer}\xspace}
\newcommand{\dsl}{\mathcal{L}_\mathrm{T}}

\definecolor{diffstart}{named}{Grey}
\definecolor{diffincl}{named}{DarkBlue}
\definecolor{diffrem}{named}{Red}

\lstdefinelanguage{diff}{
    basicstyle=\ttfamily\footnotesize,
    morecomment=[f][\color{diffstart}]{@@},
    morecomment=[f][\color{diffincl}]{+\ },
    morecomment=[f][\color{diffrem}]{-\ },
    morekeywords={def, while, return, if, foreach}
}

\lstdefinelanguage{Transformation}{
    basicstyle=\ttfamily\small,
    morekeywords={Map, Where, Update, Insert, Delete, Filter, Match, ConstNode, CNode, Type,
    RefNode, NSeq, Node, Parent}
}

\begin{document}
\bstctlcite{IEEEexample:BSTcontrol}
%
\title{Learning Syntactic Program Transformations\\from Examples}

\author{
    \IEEEauthorblockN{Reudismam Rolim\IEEEauthorrefmark{1}, Gustavo Soares\IEEEauthorrefmark{1}\IEEEauthorrefmark{2}, Loris D'Antoni\IEEEauthorrefmark{3},\\ Oleksandr Polozov\IEEEauthorrefmark{4}, Sumit Gulwani\IEEEauthorrefmark{5}, Rohit Gheyi\IEEEauthorrefmark{1}, Ryo Suzuki\IEEEauthorrefmark{6}, Bj\"{o}rn Hartmann\IEEEauthorrefmark{2}}
    \IEEEauthorblockA{\IEEEauthorrefmark{1}UFCG, Brazil, \IEEEauthorrefmark{2}UC Berkeley, USA, \IEEEauthorrefmark{3}University of Wisconsin-Madison, USA\\
    \IEEEauthorrefmark{4}University of Washington, USA, \IEEEauthorrefmark{5}Microsoft, USA, \IEEEauthorrefmark{6}University of Colorado Boulder, USA}
    \IEEEauthorblockA{reudismam@copin.ufcg.edu.br, gsoares@dsc.ufcg.edu.br, loris@cs.wisc.edu, polozov@cs.washington.edu,\\ sumitg@microsoft.com, rohit@dsc.ufcg.edu.br, ryo.suzuki@colorado.edu, bjoern@eecs.berkeley.edu}
    \vspace*{1.0cm}
}

\maketitle

\begin{abstract}

Integrated Development Environments (IDEs), such as Visual Studio, automate common transformations, such as \textit{Rename} and \textit{Extract Method} refactorings. However, extending these catalogs of transformations is complex and time-consuming.
A similar phenomenon appears in intelligent tutoring systems where instructors
have to write cumbersome code transformations that describe ``common faults''
to fix similar student submissions to programming assignments.

In this paper, we present \technique{}, a technique for automatically generating program transformations.
\technique{} builds on the observation that 
code edits performed by developers 
can be used as input-output examples for learning 
program transformations.
Example edits may share the same structure but involve different variables and subexpressions, which must be generalized in a transformation at the right level of abstraction.
To learn transformations, \technique{} leverages state-of-the-art
programming-by-example methodology using the following
key components:
(a) a novel domain-specific language (DSL) for describing program transformations,
(b) domain-specific deductive algorithms for efficiently synthesizing transformations in the DSL, and
(c) functions for ranking the synthesized transformations.

We instantiate and evaluate \technique{} in two domains.
First, given examples of code edits used
by students to fix incorrect programming assignment submissions,
we learn program transformations that can fix other students' submissions with similar
faults.
In our evaluation conducted on \totaltasks{} programming tasks performed by \totalstudents{} students, our technique helped to fix
incorrect submissions for
 \averageHelpedStudents{}\% of the students.
In the second domain, we use repetitive code edits applied by developers to the same project to synthesize a program
transformation that applies these edits to other locations in the code.
In our evaluation conducted on \numberOfRepetitiveChanges{} scenarios of repetitive edits taken from 3 large C\# open-source projects,
\technique{} learns the intended program transformation in \refactoringAccuracy{}\% of the cases and using only \examplesToEdit{} examples on average.
\end{abstract}


%
\IEEEpeerreviewmaketitle

\section{Introduction}

As software evolves, developers edit program source code to add features, fix bugs, or refactor it.
Many such \emph{edits} have already been performed in the past by the same developers in a different codebase location, or by other developers in a different program/codebase.
For instance, to apply an API update, a developer needs to locate all references to the old API and consistently replace them with the new API~\cite{ME13LASE, WA13SCAL}.
As another example, in programming courses student submissions that exhibit the same fault often need similar fixes.
For large classes such as \emph{massive open online courses} (MOOCs), manually providing feedback to different students is an unfeasible burden on the teaching staff.

Since applying repetitive edits manually is tedious and error-prone, developers often strive to automate them.
The space of tools for automation of repetitive code edits contains IDEs, static analyzers, and various domain-specific engines.
\emph{IDEs}, such as Visual Studio~\cite{visualstudio} or Eclipse~\cite{eclipse}, include features that automate some code \emph{transformations}, such as adding boilerplate code (e.g., equality comparisons or constructors) and code refactoring (e.g., \textit{Rename}, \textit{Extract Method}).
\emph{Static analyzers}, such as ReSharper~\cite{resharper}, Coverity~\cite{coverity}, ErrorProne~\cite{errorprone} and Clang-tidy~\cite{clangtidy} automate removal of suspicious code patterns, potential bugs, and verbose code fragments.
In an education context, AutoGrader~\cite{autograder} uses a set of program transformations provided by an instructor to fix common faults in introductory programming assignments.

All aforementioned tool families rely on predefined catalogs of recognized \emph{transformation} classes, which are hard to extend.
These limitations inspire a natural question:
\begin{center}
    \small
    \emph{Is it possible to learn program transformations automatically?}
\end{center}
Our key observation is that code edits gathered from repositories and version control history constitute \emph{input-output examples} for learning \emph{program transformations}.

\begin{figure}[!t]
\centering
\begin{subfigure}{0.45\textwidth}
\begin{lstlisting}[language=diff, frame = single, label=incorrect-attempt,numbers=left, numbersep=5pt, xleftmargin=.02\textwidth]
 def product(n, term):
   total, k = 1, 1
   while k<=n:
-    total = total*k
+    total = total*term(k)
     k = k+1
   return total
\end{lstlisting}
\caption{An edit applied by a student to fix the program.}
\end{subfigure}
\begin{subfigure}{0.45\textwidth}
\begin{lstlisting}[language=diff, frame = single, label=incorrect-attempt,numbers=left, numbersep=5pt, xleftmargin=.02\textwidth]
 def product(n, term):
   if (n==1):
     return 1
-  return product(n-1, term)*n
+  return product(n-1, term)*term(n)
\end{lstlisting}
\caption{An edit applied by another student fixing the same fault.}
\end{subfigure}
\begin{subfigure}{0.45\textwidth}
\centering
    \includegraphics[width=1\textwidth]{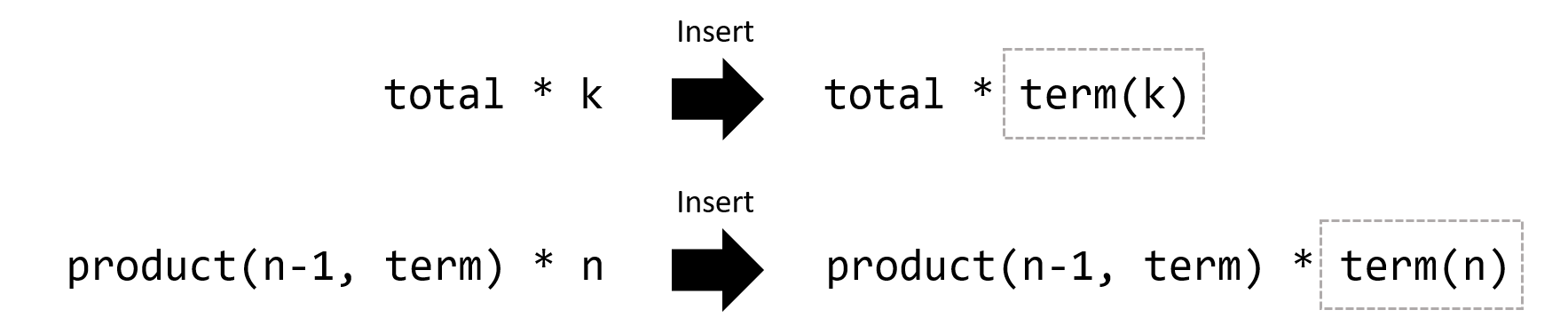}
    \caption{Similar tree edits applied in (a) and (b), respectively. Each edit inserts a concrete subtree to the right hand side of the $*$ operator. The two edits share the same structure but involve different
    variables and expressions.
    }
\end{subfigure}
\begin{subfigure}{0.45\textwidth}
\centering
    \includegraphics[width=0.9\textwidth]{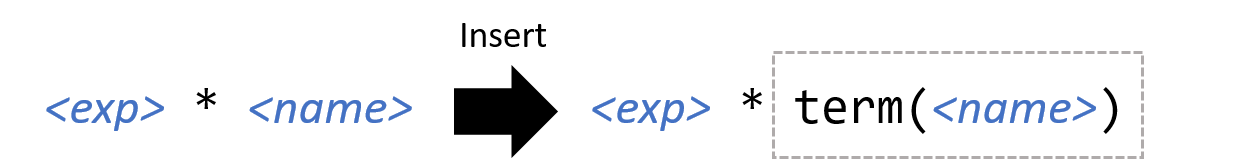}
    \caption{A rewrite rule that captures the two edits in (a) and (b).}
\end{subfigure}
\vspace{-\baselineskip}
    \caption{An example of a common fault made by different students, two similar edits that can fix different programs,
    and a program transformation that captures both edits.}
\label{fig:motivating1}
\vspace{-\baselineskip}
\end{figure}

The main challenge of example-based learning lies in abstracting concrete code edits into classes of \emph{transformations} representing these edits.
For instance, Figure~\ref{fig:motivating1} shows similar edits performed by different students to fix the same fault in their submissions for a programming assignment.
Although the edits share some structure, they involve different expressions and variables.
Therefore, a transformation should partially abstract these edits as in Figure~\ref{fig:motivating1}(d).

However, examples are highly ambiguous, and many different transformations may satisfy them.
For instance, replacing \code{<name>} by \code{<exp>} in the transformation will still satisfy the examples in Figure~\ref{fig:motivating1}.
In general, learning either the most specific or the most general transformation is undesirable, as they are likely to respectively produce false negative or false positive edits on unseen programs.
Thus, we need to \textbf{(a)}~learn and store a \emph{set} of consistent transformations efficiently, and \textbf{(b)} rank them with respect to their trade-off between over-generalization and over-specialization.
To resolve these challenges, we leverage state-of-the-art software engineering research to learn such transformations automatically using a
technique called \emph{Inductive Programming} (IP), or \emph{Programming by Examples} (PBE)~\cite{GU15INDU},
which has been successfully applied to many domains, such as text transformation~\cite{GU11AUTO}, data cleaning~\cite{LE14FLAS}, and layout transformation~\cite{edge2015mixed}.

\paraheading{Our technique}
In this paper, we propose \technique{}, an IP technique for synthesizing program transformations from examples. \technique{} is based on~\prose{}~\cite{PO15FLAS}, a state-of-the-art IP framework. We specify a \emph{domain-specific language} (DSL) that describes a rich space of program
transformations that commonly occur in practice.
In our DSL, a program transformation is defined as a sequence of distinct \emph{rewrite rules} applied to the \emph{abstract syntax tree} (AST).
Each rewrite rule matches some subtrees of the given AST and outputs modified versions of these subtrees.
Additionally, we specify constraints for our DSL operators based on the input-output examples to reduce the search space of transformations, allowing \prose{} to efficiently synthesize them. Finally, we define functions to rank the synthesized transformations based on their DSL structure.

\paraheading{Evaluation}
We evaluated \technique in two domains: learning transformations to fix submissions to introductory programming assignments and
learning transformations to apply repetitive edits to large code bases.

Our first experiment is motivated by the recent advances in \emph{massive open online courses} (MOOCs), where automatically grading student
submission and providing personalized feedback is challenging due to the large number of students.
In this experiment, we mine
existing submissions to programing assignments to collect examples of edits applied by students to fix their code.
We then use these examples to synthesize program transformations
and we try using the learned transformations to fix  any new students' submissions that exhibit similar types of faults.
We say that a submission is ``fixed'' if it passes the set of tests provided by course instructors.
In our evaluation conducted on \totaltasks{} programming tasks performed by \totalstudents{} students, \technique{} synthesizes
transformations that fix incorrect submissions for \averageHelpedStudents{}\% of the students.

Our second experiment is motivated by the fact that certain repetitive tasks occurring during software evolution, such as complex
forms of code refactoring, are beyond the capabilities of current IDEs and have to be performed manually~\cite{HI12ONTH,KN04STAT}.
In this experiment, we use repetitive code edits applied by developers to the same project to synthesize a program
transformation that can be applied to other locations in the code.
We performed a study on three popular open-source C\# projects (Roslyn~\cite{roslyn}, Entity Framework~\cite{entity}, and NuGet~\cite{nuget}) to identify and characterize repetitive code transformations.
In our evaluation conducted on \numberOfRepetitiveChanges{} scenarios of repetitive edits,
\technique{} learns the intended program transformation in \refactoringAccuracy{}\% of the cases using \examplesToEdit{} examples on average.
The learned transformations are applied to as many as \maxNumberOfLocations{} program locations.
Moreover, in~\numberOfRepetitiveChangesMoreCommit{} cases \technique synthesized transformations on more
program locations than the ones present in our dataset, thus suggesting potentially missed locations to the developers.

\paraheading{Contributions}
This paper makes the following contributions:
\begin{itemize}
    \item \technique{}, a novel technique that leverages state-of-the-art IP methodology to efficiently solve the problem of synthesizing
        program transformations from input-output examples (Section~\ref{s:technique});
    \item An evaluation of \technique{} in the context of learning fixes for students' submissions to introductory programming assignments (Section~\ref{s:education});
    \item An evaluation of \technique{} in the context of learning transformations to apply repetitive edits in open-source industrial C\# code (Section~\ref{s:refactoring}).
\end{itemize}

\section{Motivating Examples}
\label{s:motivating}

\begin{figure*}[!t]
\begin{lstlisting}[language=diff, frame = single,  numbersep=5pt, xleftmargin=.03\textwidth]
-  while (receiver.CSharpKind() == SyntaxKind.ParenthesizedExpression)
+  while (receiver.IsKind(SyntaxKind.ParenthesizedExpression))

-  foreach (var m in modifiers) {if (m.CSharpKind() == modifier) return true; };
+  foreach (var m in modifiers) {if (m.IsKind(modifier)) return true; };
\end{lstlisting}
\vspace{-\baselineskip}
\caption{Repetitive edits applied to the Roslyn source code to perform a refactoring.}
\label{fig:motivating2}
\vspace{-\baselineskip}
\end{figure*}

We start by describing two motivating examples of repetitive program transformations.

\subsection{Fixing programming assignment submissions}
\label{s:motivating1}

Introductory programming courses are often validated and graded using a test suite, provided by the instructors.
However, many students struggle to understand the fault in their code when a test fails.
To provide more detailed feedback, (e.g., fault location or its description), teachers typically compile a \emph{rubric} of common types
of faults, and detect them with simple checks.
With a large variety of possible faults, manually implementing these checks can be laborious and error-prone.

However, many faults are common and exhibit themselves in numerous unrelated student submissions.
Consider the Python code in Figure~\ref{fig:motivating1}(a).
It describes two submission attempts to solve a programming assignment in the course ``The Structure
and Interpretation of Computer Programs'' (CS61A) at UC Berkeley\footnote{\url{http://cs61a.org/}}, an introductory programming class with
more than 1,000 enrolled students.
In this assignment, the student is asked to write a porgram that computes the product
of the first \code{n} terms, where \code{term} is a function.
The original code, which includes line 4 instead of line 5, is an incorrect submission for
this assignment, and the subsequent student submission fixes it by replacing line 4 with line 5.
Notably, the fault illustrated in Figure~\ref{fig:motivating1} was
a common fault affecting more than 100 students in the Spring semester of 2016
and Figure~\ref{fig:motivating1}(b) shows a recursive algorithm proposed by a different student, which contained the same fault.

To alleviate the burden of compiling manual feedback, we propose to automatically learn the rubric checks from the student submissions.
Existing tools for such automatic learning~\cite{ME11SYST,ME13LASE} cannot generate a transformation that is general enough to represent
both the edits shown in Figure~\ref{fig:motivating1}(c) due to their limited forms of abstraction.
In \technique{}, this transformation is described as a rewrite rule shown in Figure~\ref{fig:motivating1}(d).
This rewrite rule pattern matches any subtree of the program's AST whose root is a \code{*} operation with a variable as the second operand,
and inserts a \code{term} application on top of that variable.
Notice that the rewrite rule abstracts both a variable name and the first operand of the \code{*} operator.

\subsection{Repetitive codebase edits}
\label{s:motivating2}

We now illustrate how \technique{} automates repetitive codebase editing.
The following example is found in Roslyn, the Microsoft's library for compilation and code analysis for C\# and VB.NET~\cite{roslyn}.
Consider the edits shown in Figure~\ref{fig:motivating2}, where,
for every instance of a comparison with an object returned by the method \code{CSharpKind},
the developer replaces the \code{==} operator with an invocation of the new method \code{IsKind}, and
passes the right-hand side expression as the method's argument.
Such refactoring is beyond the abilities of existing IDEs due to its context sensitivity.
In contrast, \technique{} generalizes the two example edits in Figure~\ref{fig:motivating2} to
the intended program transformation, which can then be
applied to all other matching AST subtrees in the code.

When we analyzed the  commit \textit{8c14644}\footnote{\url{https://github.com/dotnet/roslyn/commit/8c14644}} in the Roslyn repository, we observed that the developer
applied this edit to 26 locations in the source code.
However, the transformation generated by \technique{} applied this edit to 689 more locations.
After we presented the results to the Roslyn developers, they confirmed that the locations discovered by \technique should have been covered
in the original commit.

\section{Technique}
\label{s:technique}

In this section, we describe our technique for synthesizing program transformations from input-output examples.
\technique builds on \prose{}~\cite{PO15FLAS}, a framework for program synthesis from examples and under-specifications.

In \prose, an application designer defines a \emph{domain-specific language} (DSL) for the desired tasks.
The synthesis problem is given by a \emph{spec} $\varphi$, which contains a set of program inputs and constraints on the desired program's outputs on
these inputs (e.g., examples of these outputs).
\prose synthesizes a set of programs in the DSL that is consistent with $\varphi$, using a combination of \emph{deduction}, \emph{search},
and \emph{ranking}:
\begin{itemize}
    \item Deduction is a top-down walk over the DSL grammar, which iteratively \emph{backpropagates} the spec $\varphi$ on the desired
        program to necessary specs on the subexpressions of this program.
        In other words, it reduces the synthesis problem to smaller synthesis subproblems using a divide-and-conquer dynamic programming
        algorithm over the desired program's structure.
    \item Search is an enumerative algorithm, which iteratively constructs candidate subexpressions in the grammar and verifies them for
        compliance with the spec $\varphi$~\cite{sygus}.
    \item Ranking is a process of picking the most robust program from the synthesized set of programs that are consistent with $\varphi$.
        Because examples are highly ambiguous, such a set may contain up to $10^{20}$ programs~\cite{PO15FLAS}, and quickly
        eliminating undesirable candidates is paramount for a user-friendly experience.
\end{itemize}

\begin{figure*}
    \centering
    \includegraphics[width=0.68\textwidth]{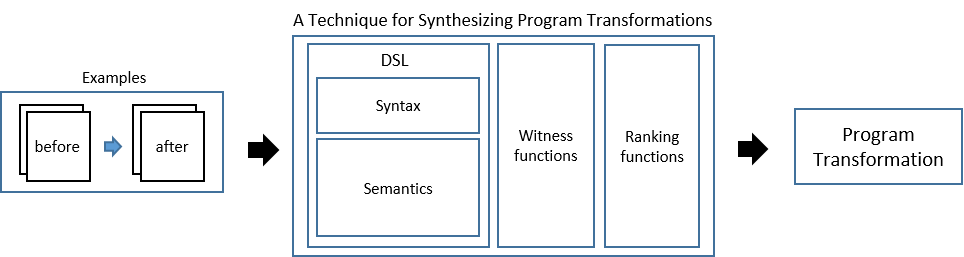}
    \caption{The workflow of \technique.
    It receives an example-based specification of edits as input, and returns a transformation.}
    \label{fig:process}
    \vspace{-\baselineskip}
\end{figure*}

\technique consists of three main components, which are illustrated in Figure~\ref{fig:process}:

\begin{itemize}
    \item \textit{A DSL for describing program transformations}. It contains operators that allow partially abstracting edits provided as examples. The DSL is expressive enough for representing common transformations but restrict enough to allow efficient synthesis.  
    \item \textit{Witness functions}.
        In \prose, a \emph{witness function} $\omega_F$ is a backpropagation procedure, which, given a spec $\varphi$ on a desired
        program on kind $F(e)$, deduces a necessary (or even sufficient) spec $\varphi_e = \omega_F(\varphi)$ on its
        subexpression~$e$.~\footnote{Another view on witness functions $\omega_F$ is that they simply implement \emph{inverse semantics} of
        $F$, or a generalization of inverse semantics w.r.t. some \emph{constraints} on the output of $F$ instead of just its \emph{value}.}
        Witness functions enable efficient top-down synthesis algorithms of \prose.
    \item \textit{Ranking functions}.
        Since example-based specifications are incomplete, the synthesized abstract transformation may not perform the desired transformation on
        other input programs.
        We specify \emph{ranking functions} that rank a transformation based on its robustness (i.e., likelihood of it being correct in general).
\end{itemize}

\subsection{A DSL of AST transformations}
\label{s:core}

\begin{figure}
    \small
    \setlength{\grammarindent}{5.9em} 
    \begin{grammar}
        <transformation> ::=  "Transformation("<rule>$_1$, \dots, <rule>$_n$")"

        <rule> ::= "Map("$\lambda x \rightarrow$ <operation>, <locations>")"

        <locations> ::= "Filter("$\lambda x \rightarrow$ "Match("$x$", "<match>")", "AllNodes())"

        <match> ::= "Context("<pattern>", "<path>")"

        <pattern> ::= <token> | "Pattern("<token>", "<pattern>$_1$, \ldots, <pattern>$_n$")"

        <token> ::= "Concrete("$kind$"," $value$")" | "Abstract("$kind$")"

        <path> ::= "Absolute("$s$")" | "Relative("<token>", "$k$")"

        <operation> ::= "Insert("$x$"," <ast>"," $k$")" \alt "Delete("$x$"," <ref>")" \alt "Update("$x$"," <ast>")" \alt "InsertBefore("$x$","<ast>")"

        <ast> ::= <const> | <ref>

        <const> ::= "ConstNode("$kind$"," $value$"," <ast>$_1$, \ldots, <ast>$_n$")"

        <ref> ::= "Reference("$x$"," <match>"," $k$")"

    \end{grammar}
    \caption{A core DSL $\dsl$ for describing AST transformations.
        $kind$ ranges over possible AST kinds of the underlying programming language, and $value$ ranges over all possible ASTs.
        $s$~and~$k$ range over strings and integers, respectively.}
    \label{fig:dsl}
\end{figure}

In this section, we present our DSL of program transformations, hereinafter denoted $\dsl$.
It is based on tree edit operators (e.g., \code{Insert}, \code{Delete}, \code{Update}), list processing operators (\code{Filter},
\code{Map}), and pattern-matching operators on trees.
The syntax of $\dsl$ is formally given in Figure~\ref{fig:dsl}.

A \emph{transformation} $T$ on an AST is a list of \emph{rewrite rules} (or simply ``rules'') $r_1$, \ldots, $r_n$.
Each rule $r_i$ specifies an \emph{operation}~$O_i$ that should be applied to some set of \emph{locations} in the input AST.
The locations are chosen by \emph{filtering} all nodes within the input AST w.r.t. a \emph{pattern-matching predicate}.

Given an input AST $P$, each rewrite rule $r$ produces a \emph{list of concrete edits} that may be applied to the AST.
Each such edit is a replacement of some node in $P$ with a new node.
This set of edits is typically an overapproximation of the desired transformation result on the AST; the precise method for applying the
edits is domain-specific (e.g., based on verification via unit testing).
We discuss the application procedures for our studied domains in Section~\ref{s:evaluation}.
In the rest of this subsection, we focus on the semantics of rewrite rules, which produce the suggested edits.

A rewrite rule consists of two parts: a \emph{location expression} and an \emph{operation}.
A location expression is a \code{Filter} operator on a set of sub-nodes of a given AST.
Its predicate $\lambda\, x \to$ \code{Match}($x$, \code{Context}($pattern$, $path$)) matches each sub-node~$x$ with a \emph{pattern expression}.

\paraheading{Patterns}
A pattern expression \code{Context}($pattern$, $path$) checks the \emph{context} of the node $x$ against a given $pattern$.
Here $pattern$ is a combination of \code{Concrete} tokens (which match a concrete AST) and \code{Abstract} tokens (which match only the AST kind).
In addition, a \emph{path expression} specifies the expected position of $x$ in the context that is described by $pattern$, using an
notation similar to XPath~\cite{xpath}.
This allows for a rich variety of possible pattern-matching expressions, constraining the ancestors or the descendants of the desired
locations in the input AST.

\begin{figure}[t]
\begin{subfigure}{0.5\textwidth}
\includegraphics[width=1\textwidth]{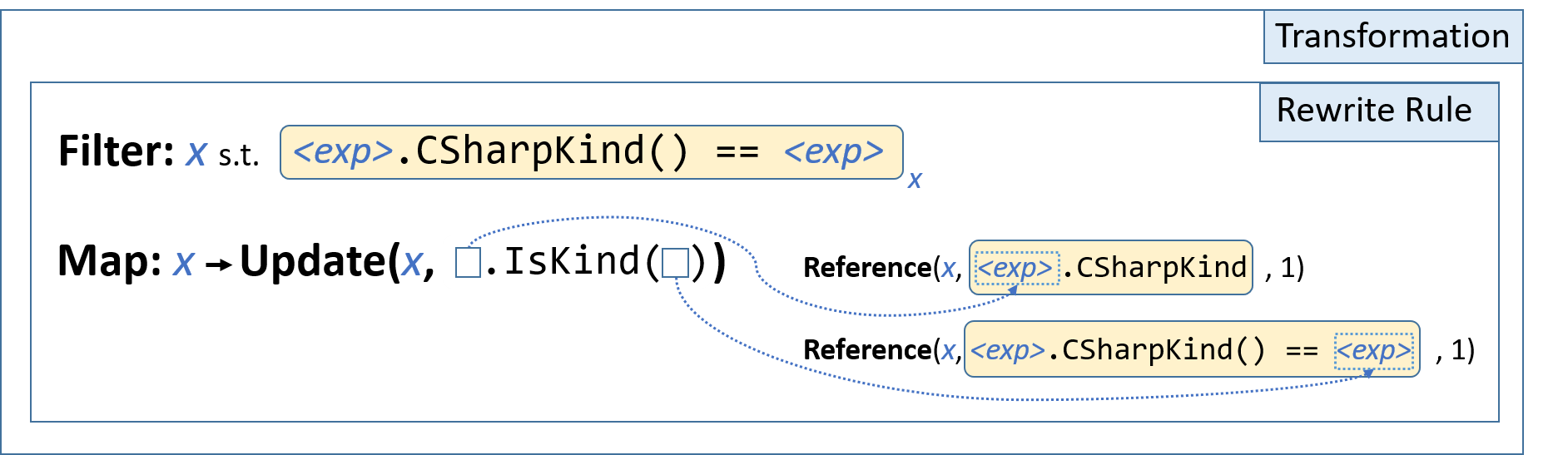}
\caption{A synthesized AST transformation.}
\end{subfigure}
\begin{subfigure}{0.48\textwidth}
\begin{lstlisting}[language=diff, frame = single,  numbersep=5pt, xleftmargin=.02\textwidth]
while (receiver.CSharpKind() ==
  SyntaxKind.ParenthesizedExpression) {
  ...
}

foreach (var m in modifiers) {
  if (m.CSharpKind() == modifier)
    return true;
};
...
if (r.Parent.CSharpKind() ==
  SyntaxKind.WhileStatement) {
  ...
}
\end{lstlisting}
\caption{A C\# program used as an input to the transformation.}
\end{subfigure}
\begin{subfigure}{0.48\textwidth}
\includegraphics[width=1\textwidth]{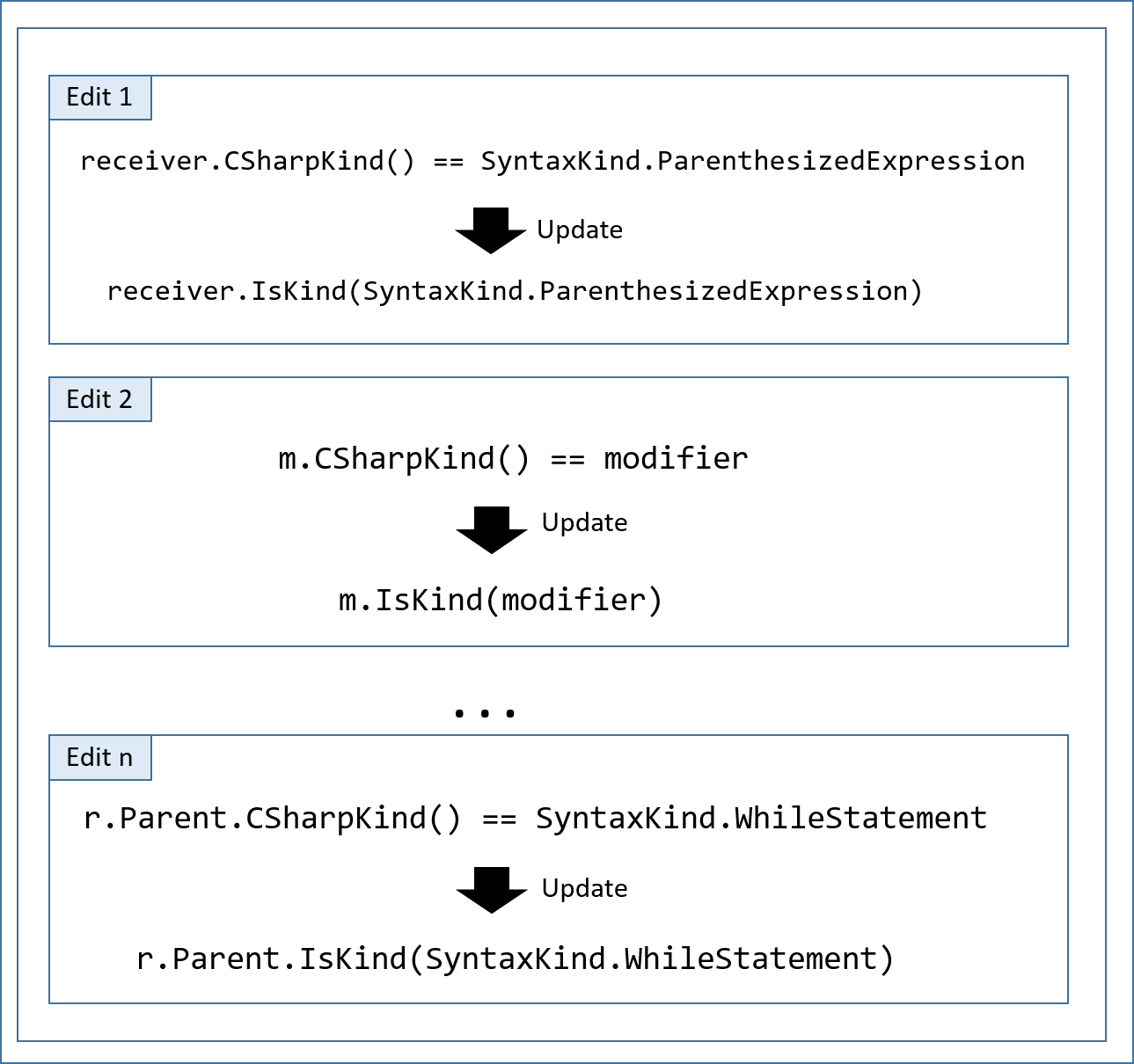}
\caption{A list of edits produced after instantiating (a) to (b).}
\end{subfigure}
\caption{An example of a synthesized transformation and its application to a C\# program, which results in a list of edits.}
\vspace{-\baselineskip}
\label{fig:transformation}
\end{figure}

\begin{example}
    Figure~\ref{fig:transformation} shows a transformation that describes our running example from Figure~\ref{fig:motivating2}.
    This transformation contains one rewrite rule.
    Its location expression is
    \begin{center}
        \code{Filter}($\lambda x \to$ \code{Context}($\pi$, \code{Absolute}(\code{""})))
    \end{center}
    where
    \begin{align*}
        \pi &= \text{\code{Pattern}}(\kind{==}, \text{\code{Pattern}}(\kind{.}, t_e, t_m), t_e) \\
        t_e &= \text{\code{Abstract}}(\kind{<exp>}) \\
        t_m &= \text{\code{Concrete}}(\kind{<call>}, \text{\code{"CSharpKind()"}})
    \end{align*}
    The path expression \code{Absolute("")} specifies that the expected position of a location $x$ in $\pi$ should be at the root -- that
    is, the pattern $\pi$ should match the node $x$ itself.
    \label{ex:filter}
\end{example}

\paraheading{Operations}
Given a list of locations selected by the \code{Filter} operator, a rewrite rule applies an \emph{operation} to each of them.
An operation $O$ takes as input an AST $x$ and performs one of the standard tree edit procedures~\cite{Pawlik:2011:RRA:2095686.2095692,zhang-sasha} on it:
\begin{itemize}
    \item \code{Insert} some fresh AST as the $k^{\text{th}}$ child of $x$
    \item \code{Delete} some sub-node from $x$
    \item \code{Update} $x$ with some fresh AST
    \item \code{InsertBefore}: insert some fresh AST as the preceding sibling of $x$
\end{itemize}

An operation creates fresh ASTs using a combination of \emph{constant ASTs} \code{ConstNode} and \emph{reference ASTs} \code{Reference},
extracted from the location node $x$.
Reference extraction uses the same pattern-matching language, described above.
In particular, it can match over the ancestors or descendants of the desired reference.
Thus, the semantics of reference extraction \code{Reference}($x$, \code{Context}($pattern$, $path$), $k$) is:
\begin{enumerate}
    \item Find all nodes in $x$ s.t. their surrounding context matches $pattern$, and they are located at $path$ within that
        context.
    \item Out of all such nodes, extract the $k^{\text{th}}$ one.
\end{enumerate}

\begin{example}
    For our running example from Figure~\ref{fig:motivating2}, the desired rewrite rule applies the following operation to all nodes
    selected by the location expression from Example~\ref{ex:filter}:
    \begin{center}
        \code{Update}($x$, \code{ConstNode}(\kind{.}, $\ell_1$, \kind{<call>}, \code{"IsKind"}, $\ell_2$))
    \end{center}
    where
    \begin{align*}
        \ell_1 &= \text{\code{Reference}}(x, \text{\code{Context}}(\pi_1, s_1), 1) \\
        \ell_2 &= \text{\code{Reference}}(x, \text{\code{Context}}(\pi_2, s_2), 1) \\
        \pi_1 &= \text{\code{Pattern}}(\kind{.}, t_e, t_m) \\
        \pi_2 &= \text{\code{Pattern}}(\kind{==}, \text{\code{Pattern}}(\kind{.}, t_e, t_m), t_e) \\
        s_1 &= \text{\code{Absolute("1")}} \qquad s_2 = \text{\code{Absolute("2")}}
    \end{align*}
    and $t_e$ and $t_m$ are defined in Example~\ref{ex:filter}.
    This operation updates the selected location $x$ with a fresh call to \code{IsKind}, performed on the extracted receiver AST from $x$,
    and with the extracted right-hand side AST from $x$ as its argument.
\end{example}

\subsection{Synthesis algorithm}

We now describe our algorithm for synthesizing AST transformations from input-output examples.
Formally, it solves the following problem: given an example-based spec $\varphi$, find a transformation $T \in \dsl$ that is consistent with
all examples \mbox{$(P_i, P_o) \in \varphi$}.
We denote this problem as $T \vDash \varphi$.

Recall that the core methodology of PBE in \prose is \emph{deductive synthesis}, or \emph{backpropagation}.
In it, a problem of kind $F(T_1, T_2) \vDash \varphi$ is reduced to several subproblems of kinds $T_1 \vDash \varphi_1$ and $T_2 \vDash
\varphi_2$, which are then solved recursively.
Here $\varphi_1$ and $\varphi_2$ are fresh specs, which constitute necessary (or even sufficient) constraints on the subexpressions
$T_1$ and $T_2$ in order for the entire expression $F(T_1, T_2)$ to satisfy $\varphi$.
In other words, the examples on an operator $F$ are backpropagated to examples on the parameters of $F$.

As discussed previously, the backpropagation algorithm relies on a number of modular operator-specific annotations called \emph{witness
functions}.
Even though \prose includes many generic operators with universal witness functions out of the box (e.g. list-processing \code{Filter}),
most operators in $\dsl$ are domain-specific, and therefore require non-trivial domain-specific insight to enable backpropagation.
The key part of this process is the witness function for the top-level \code{Transformation} operator.

The operator \code{Transformation}($rule_1$, \dots, $rule_n$) takes as input a list of rewrite rules and produces a transformation
that, on a given AST, applies these rewrite rules in all applicable locations, producing a list of edits.
The backpropagation problem for it is stated in reverse: given examples $\varphi$ of edits performed on a given AST, find necessary
constraints on the rewrite rules $rule_1$, \dots, $rule_n$ in the desired transformation.

The main challenges that lie in backpropagation for \code{Transformation} are:
\begin{enumerate}
    \item Given an input-output example $(P_i, P_o)$, which often represents the entire codebase/namespace/class, find examples of
        individual edits in the AST of $P_i$.
    \item Partition the edits into clusters, deducing which of them were obtained by applying the same rewrite rule.
    \item For each cluster, build a set of operation examples for the corresponding rewrite rule.
\end{enumerate}

\paraheading{Finding individual edits}
We resolve challenge 1 by calculating \emph{tree edit distance} between $P_i$ and $P_o$.
Note that the state-of-the-art Zhang-Shasha tree edit distance algorithm~\cite{zhang-sasha} manipulates single nodes, whereas our operations
(and, consequently, examples of their behavior) manipulate whole subtrees.
Thus, to construct proper examples for operations in $\dsl$, we group tree edits computed by the distance algorithm into connected
components.
A connected component of node edits represents a single edit operation over a subtree.
\loris{not sure I get what this last sentence means}

\paraheading{Partitioning into rewrite rules}
To identify subtree edits that were performed by the same rewrite rule, we use the DBSCAN~\cite{dbscan} clustering algorithm to partition
edits by similarity.
Here we conjecture that components with similar edit distances constitute examples of the same rewrite rule.

Algorithm~\ref{algo:witness} describes the steps performed by the witness function for \code{Transformation}.
Lines 2-6 perform the steps described above: computing tree edit distance and clustering the connected components of edits.
Then, in lines 7-11, for each similar component, we extract the topmost operation to create an example for the corresponding rewrite rule.
This example contains the subtree where the operation was applied in the input AST and the resulting subtree in the output AST.

\begin{algorithm}[t]
    \small
    \caption{Backpropagation procedure for the DSL operator \code{Transformation}($rule_1$, \dots, $rule_n$).}
    \label{algo:repeat}
    \begin{algorithmic}[1]
        \REQUIRE{Example-based spec $\varphi$}
        \STATE{$result$ $\leftarrow$ dictionary for storing examples for each input}
        \FORALL {($P_i$,$P_o$) in $\varphi$}
        \STATE{$examples$ $\leftarrow$ empty list of refined  examples for edits}
        \STATE{$operations$ $\leftarrow$ \textsc{TreeEditDistance}($P_i$, $P_o$)}
        \STATE{$components$ $\leftarrow$ \textsc{ConnectedComponents}($operations$)}
        \STATE{$connectedOpsByEdits$ $\leftarrow$ \textsc{DBScan}($components$)}
        \FORALL {$connectedOps \in connectedOpsByEdits$}
        \STATE{$ruleExamples \rightarrow connectedOps$.\textsc{Map}( \\ \quad $ops\rightarrow$ create a single concrete operation based on $ops$)}
        \STATE{$examples$.Add($ruleExamples$)}
    \ENDFOR
    \STATE{$result$[$P_i$].Add($examples$)}
\ENDFOR
\RETURN{$result$}
\end{algorithmic}
\label{algo:witness}
\end{algorithm}




\subsection{Ranking}

The last component of \technique is a ranking function for transformations synthesized by the backpropagation algorithm.
Since $\dsl$ typically contains many thousands of ambiguous programs that are all consistent with a given example-based spec, we must
disambiguate among them.
Our ranking function selects a transformation that is more likely to be robust on unseen ASTs -- that is, avoid false positive and false
negative matches.
It is based on the following principles:

\begin{itemize}
    \item Favor \code{Reference} over \code{ConstNode}: a transformation that resues a node from the input AST is more likely to satisfy the
        intent than a transformation that constructs a constant AST.
    \item Favor patterns with non-root paths, that is patterns that consider surrounding context of a location.
        A transformation that selects its locations based on surrounding context is less likely to generate false positives.
    \item Among patterns with non-empty context, favor the shorter ones.
        Even though context helps prevent underfitting (i.e., false positive matches), over-specializing to large contexts may lead to
        overfitting (i.e., false negative matches).
\end{itemize}

\section{Evaluation}
\label{s:evaluation}

In this section, we present two empirical studies to evaluate \technique{}, our technique for learning program transformations. First, we present an empirical study on learning transformations for fixing student submissions to introductory Python programming assignments (Section~\ref{s:education}). Then, we present an evaluation of \technique{} on learning transformations to apply repetitive edits to open-source C\# projects (Section~\ref{s:refactoring}). The experiments were performed on a PC Core i7 and 16GB of RAM, running Windows 10 x64 with .NET Framework 4.6.

\subsection{Fixing introductory programming assignments}
\label{s:education}
\gustavo{I am replacing mistake and error by fault, which is the common term used by SE community. Also, replacing attempts by submissions}
In this study, we use \technique{}
to learn transformations that describe how students modify an incorrect piece
of code to obtain  a correct one. We then measure how often the learned transformations can
be used to fix the incorrect code submitted by other students.  Transformations that can be applied across students are valuable because they can be used to automatically generate hints to students on how to fix bugs in their code; alternatively, they can also help TAs with writing better manual feedback. We focus our evaluation on the transfer of transformations, and leave the evaluation of hint generation to future work.

Our goal is to investigate both the overall effectiveness of our technique, and to what extent learned transformations in an education scenario are problem-specific, or general in nature. If most transformations are general purpose, instructors might be able to provide them manually, once. However, if most transformations are problem-specific, automated techniques such as \technique{} will be especially valuable.
Concretely, we address the following research questions:
\begin{description}
    \item[RQ1]
    How often can transformations learned from student code edits
    be used to fix incorrect code of other students
    who are solving the \emph{same} programming assignment?
    \item[RQ2] How often can transformations learned from student code edits
    be used to fix incorrect code of other students
    who are solving a \emph{different} programming assignment?
\end{description}

\paraheading{Benchmark}
We collected data from the introductory programming course CS61A at UC Berkeley. More than 1,000 students enroll in this course every semester, which has led the instructors to adopt solutions common to MOOCs such as video lessons and autograders. For each homework problem, the teachers provide a black-box test suite and the students
use these tests to check the correctness of their programs. The system logs
a submission whenever the student runs the provided test suite for a homework assignment.
This log thus provides a history of all submissions.
Our benchmark comprises \totaltasks{} assigned problems (see Table~\ref{tab:benchmark1}).
For each problem, students had to implement a single function in Python. We filtered the log data to focus on students who had at least one incorrect submission, which is required to learn a transformation from incorrect to correct state. We analyzed \totalattempts{} incorrect submissions, from up to \totalstudents{} students.


\paraheading{Experimental setup}
For each problem, each student in the data set submitted one or more incorrect submissions and, eventually, a correct submission.
We used the last incorrect submission and the correct submission as input-output examples
to synthesize a program transformation and used the synthesized
transformation to attempt fixing other student
submissions.
By selecting a pair of incorrect and correct submissions, we learn a transformation that changes the state of the program from incorrect to correct, fixing existing faults in the code. Students may have applied additional edits, such as refactorings, though. The transformation thus may contain unnecessary rules to fix the code. By learning from the last incorrect submission, we increase the likelihood of learning a transformation that is focused on fixing the existing faults. We leave for future work the evaluation of learning larger transformations from earlier incorrect submissions to correct submissions, and how these transformations can help fixing larger conceptual faults in the code.
We used the teacher-provided test suites to check whether a program was fixed.

For our first research question, we considered two different scenarios: \emph{Batch} and \emph{Incremental}.
In the \emph{Batch} scenario, for each programming assignment,
we synthesize transformations for all but one students in the data set and use them to fix the incorrect submissions of the single remaining
student, in a leave-one-out cross-validation -- i.e., we attempt fixing the submission of a student using only transformations learned using
submissions of other students.
This scenario simulates the situation in which instructors have data from one or more previous semesters.
In the \emph{Incremental} scenario, we sort our data set by submission time and try to fix a submission using only transformations learned from earlier timestamps. This scenario simulates the situation
in which instructors do not have previous data. Here the effectiveness of the technique increases over time.
For our second research question, we use all transformations learned in \emph{Batch} from one assignment
to attempt fixing the submissions for a different assignment.

In general, each synthesized rule in the transformation may be applicable to many locations in the code.
In our experiments, we try to apply each synthesized transformation to at most 500 combinations of locations.
If the transformation can be applied to further locations, we simply abort and proceed to the next
program transformation.


\begin{table}
    \centering
    \caption{Our benchmarks and incorrect student submissions.}
    \begin{tabular}{clcc}
        \toprule
        \multicolumn{2}{c}{Assignment} & Students & Incorrect submissions  \\
        \midrule
        Product & product of the first $n$ terms & 549 & 6,410 \\
        Accumulate & fold-left of the first $n$ terms & 668 & 6,410 \\
        Repeated & function composition, depth $n$ & 720 & 9,924 \\
        G & $G(n) = \sum_{i=1}^3 i \cdot G(n - i)$ & 379 & 2,229 \\
        \bottomrule
    \end{tabular}
    \label{tab:benchmark1}
\end{table}

\paraheading{Learned transformations are useful within the same programming assignments}
In the {\em Batch} scenario, \technique{} generated fixes for \averageHelpedStudents{}\% of the students.
While, on average, students took \averageAttemptsToComplete{} submissions to finish the assignment, the
transformations learned using \technique{}
fixed the student submissions after an average of \averageAttemptsToFix{} submissions.
In the {\em Incremental} scenario, \technique{} generated fixes for \averageHelpedStudentsIncremental{}\% of the students and required,
on average, \averageAttemptsToFixIncremental{} submissions to find a fix.
The results suggest that the technique can be useful even in the absence of data from previous semesters but using existing data can double its effectiveness.
Table~\ref{tab:resultsMistakes1} summarizes the results for the Batch and the Incremental scenarios.

\begin{figure}
    \centering
    \begin{subfigure}[b]{0.5\textwidth}
        \includegraphics[width=1\textwidth]{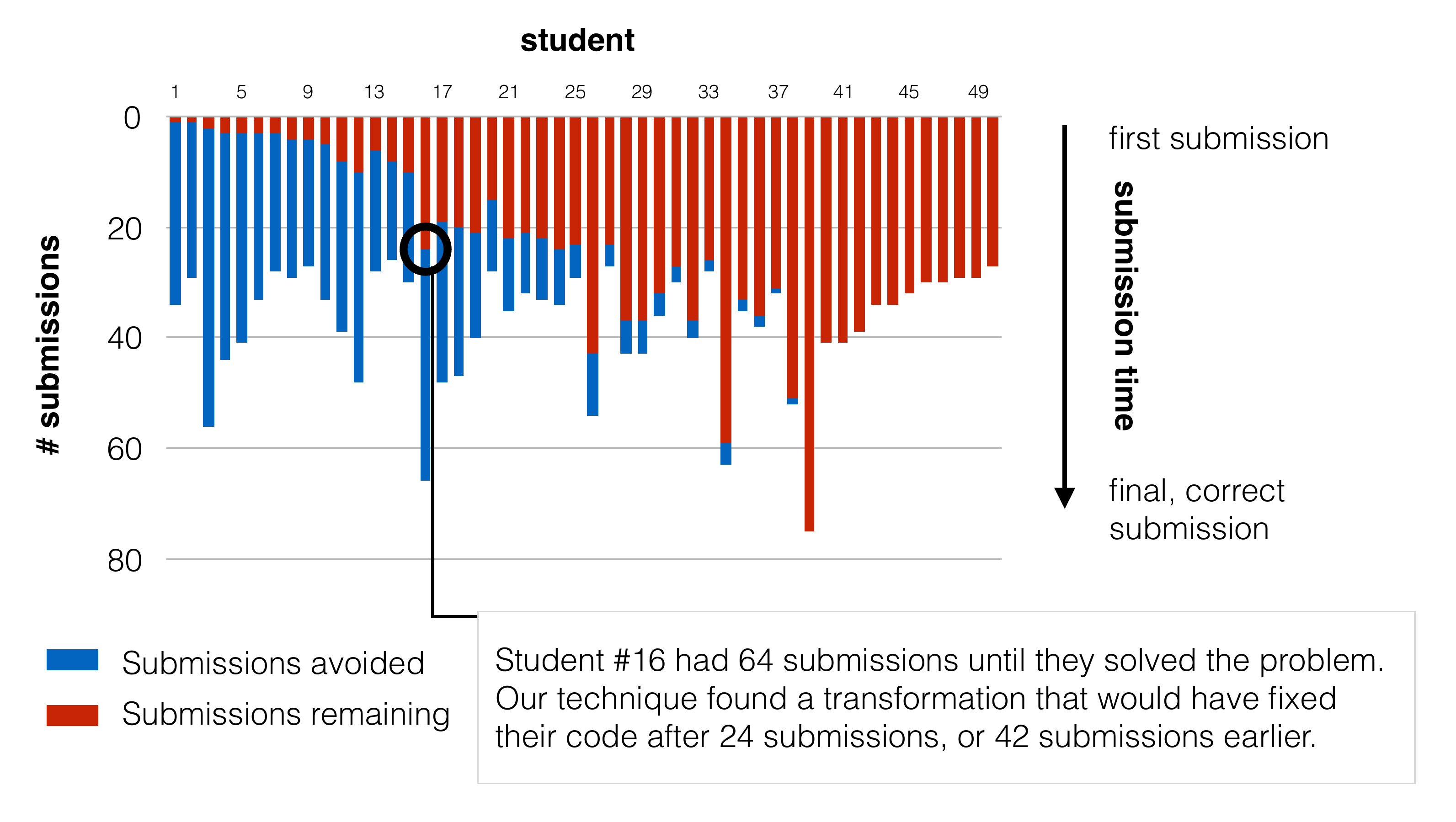}
        \caption{Assignment ``Accumulate''}
        \label{fig:subim2}
    \end{subfigure}
    \begin{subfigure}[b]{0.5\textwidth}
        \includegraphics[width=1\textwidth]{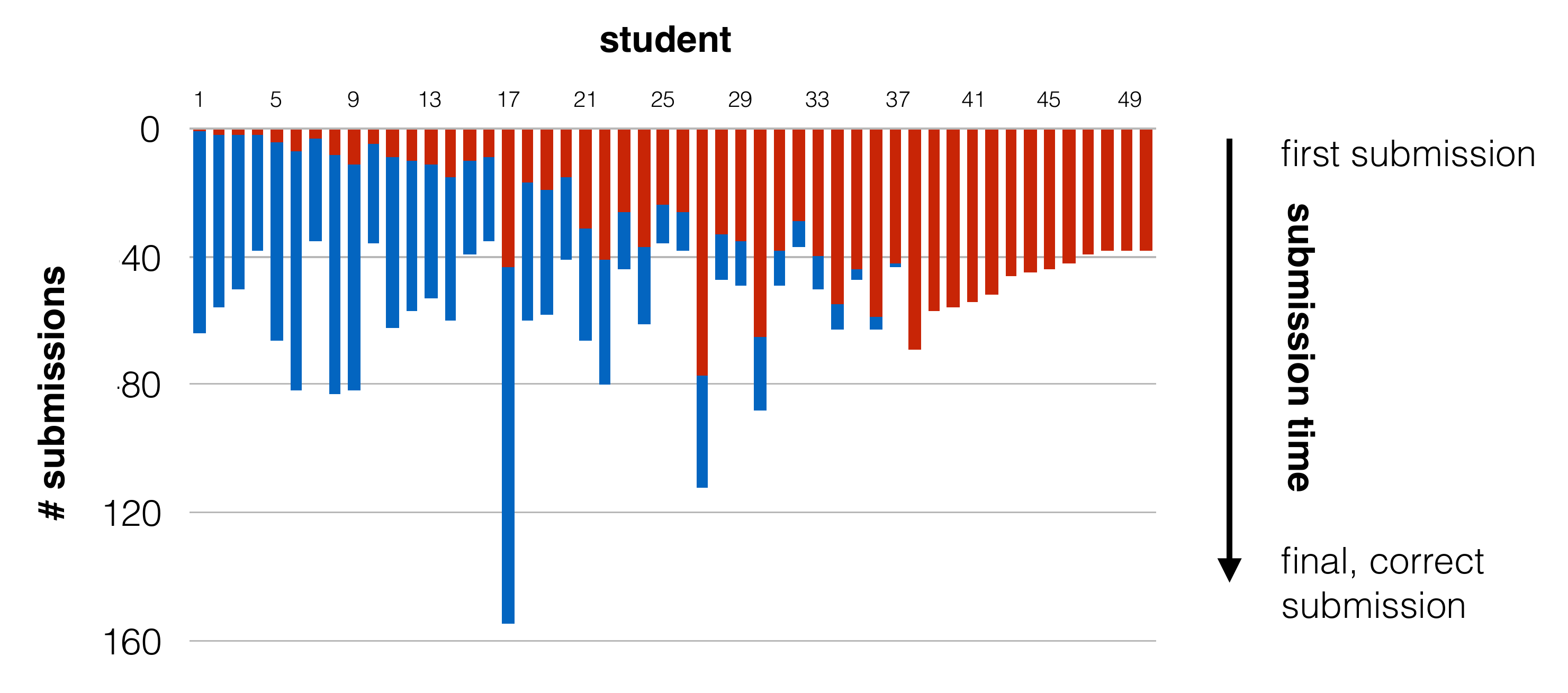}
        \caption{Assignment ``Repeated''}
        \label{fig:subim2}
    \end{subfigure}
    \caption{Analysis of the first time \technique{} can fix a student submission for the 50 students with the highest number of attempts in two benchmark problems. Blue are submissions that might be avoided by showing appropriate feedback from a fix generated by \technique{}.}
    \label{fig:top50students}
    \vspace{-\baselineskip}
\end{figure}

Although we only used students' last incorrect submissions together with their corresponding correct submissions as examples for synthesizing transformations, we could find a transformation to fix student solutions 3.5 submissions before the last incorrect submission, on average.
This result suggests that \technique{} can be used to provide feedback to help students before they know how to arrive at a correct solution themselves.
Additionally, providing feedback about mistakes can be more important for students who struggle with the assignments.
Figure~\ref{fig:top50students} shows the 50 students who had the most submission for the two hardest assignments in our benchmark.
Each column shows chronological submissions for one student, with the earliest submissions at the top, and the eventual correct submission at the bottom.
Red indicates an incorrect submission; blue shows the first time \technique{} was able to automatically fix the student's code (we only show the earliest time and do not re-test subsequent incorrect submissions).
As we can see in the charts, students took dozens (up to 148) submissions.
In many cases, our technique already provided a fix after the student attempted half of the submissions.

The transformations learned by \technique{} contain edits with different granularity, ranging from edits to single nodes in the AST, e.g., updating a constant, to edits that add multiple statements, such as adding a base case, a return statement, or even replacing an iterative solution by a recursive one. We also noticed transformations containing multiple rules that represent multiple mistakes in the code. \loris{maybe we can measure
the granularity and give a table?}

\begin{table}
\setlength{\tabcolsep}{4pt}
    \centering
    \caption{Summary of results for RQ1. ``Submissions'' = mean (SD) of incorrect submissions per student; ``students''~= \% of students that got their solution fixed by \technique{}; ``fixed''~= mean (SD) of submissions required to find the fix.}
    \begin{tabular}{cccccc}
        \toprule
        \multirow{2}{5em}{Assignment}  &  \multirow{2}{4.8em}{Submissions} & \multicolumn{2}{c}{Batch} & \multicolumn{2}{c}{Incremental} \\
        \cmidrule(lr){3-4}
        \cmidrule(l){5-6}
        & & students & fixed & students & fixed \\
        \midrule
         Product & 5.3 (8.2) & 501 (91\%) & 3.57 (6.1) & 247 (45\%) & 4.1 (6.7) \\
         Accumulate & 8.9 (10.5) & 608 (91\%) & 5.4 (7.9) & 253 (38\%) & 7.5 (9.8) \\
         Repeated & 12.7 (15.3) & 580 (81\%) & 8 (10.3) & 340 (47\%) & 9.6 (11.5) \\
         G & 5.5 (9.4) & 319 (84\%) & 1.4 (1.7) & 174 (46\%) & 4.1 (7) \\
         \midrule
         Total & 8.7 (12) & 2,008 (87\%) & 5.2 (8.1) & 1,014 (44\%) & 6.8 (9.7)\\
         \bottomrule
    \end{tabular}
    \label{tab:resultsMistakes1}
    \vspace{-\baselineskip}
\end{table}

\paraheading{Learned transformations are not useful among different programming assignments} Using the  transformations learned from other assignments, we were able to fix solutions for only 7--9\% of the students, which suggests that most transformations are problem-specific and not common among different assignments (see Table~\ref{tab:resultsMistakes2}).
The results suggest that different
assignments exhibit different fault patterns; therefore, problem-specific training corpora are needed.
This finding also suggest that other automatic grading tools that use a fixed or user-provided fault rubric (e.g., AutoGrader~\cite{autograder}) are not likely to work on arbitrary types of assignments.

\begin{table}
    \centering
    \footnotesize
    \caption{Summary of results of re-using  transformations on different assignments.}
    \begin{tabular}{ccc}
        \toprule
    Original Assignment & Target Assignment & Helped students \\
    \midrule
    Product     & G & 28 (7\%)  \\
    Accumulate     &  G & 33 (9\%) \\
    \bottomrule
    \end{tabular}
    \label{tab:resultsMistakes2}
\end{table}

\paraheading{Qualitative Feedback from Teaching Assistant}
To validate the quality of the learned transformations, we built a user interface that allows one to explore, for each transformation, the incorrect submissions that can be fixed with it. We asked a Teaching Assistant (TA) of the CS61a course to analyze the fixes found using \technique{}. The TA confirmed that fixes were generally appropriate, but also reported some issues. First, a single syntactic transformation may represent multiple distinct mistakes. For instance, a transformation that changes a literal to \code{1} was related to a bug in the stopping condition of a while loop in one student's code; and also to a bug in the initial value of a multiplication which would always result in 0 in another student's code. In this case, the TA found it hard to provide a meaningful description of the fault beyond ``replace 0 to 1''. If fixes are used to generate feedback, TAs will need additional tools to merge or split clusters of student submissions.


Finally, our technique relies on test cases for evaluating the correctness of fixed programs. Although some fixed programs passed the test cases, when test cases were incomplete, some faults remained. While reliance on test cases is a fundamental limitation, when fixes are reviewed in an interactive setting as with our TA, our technique can be used to discover the need for more test cases for particular assignments.

\subsection{Applying repetitive edits to open-source C\# projects}
\label{s:refactoring}
In this study, we use \technique{} to learn transformations that describe simple edits that have to be applied to many locations of a C\# code base. We then measure how often the learned transformation is the intended one and whether it is correctly applied to all the required code locations.
Concretely, we address the following question:

\begin{description}
    \item[RQ3]
    Can \technique{} synthesize transformations with repetitive edits to large open-source projects?
\end{description}

\paraheading{Benchmark} We manually inspected \totalNumberOfCommits{} commits from three large open source projects: Roslyn, Entity Framework, and NuGet. The projects' size range from 150,000 to 1,500,000 lines of code. We consider an edit to be repetitive if it is applied to more than two locations in the codebase. We identified~\numberOfRepetitiveChanges{} distinct scenarios of repetitive edits:~\numberOfChangesRoslyn{} in Roslyn,~\numberOfChangesEntityFramework{} in Entity Framework, and~\numberOfChangesNuGet{} in NuGet. The number of edited locations in each scenario ranges from 3 to 60, with a median of~\medianPerChange{}. Each project contains at least one scenario with more than 19 edited locations.
In \numberOfMultipleFiles{} (\percentOfMultipleFiles{}\%) out of the \numberOfRepetitiveChanges{} scenarios, there are edited locations in more than one file, which are harder to handle correctly for developers. Finally, in \numberOfNonIdentical{} (\percentOfNonIdentical{}\%) out of the~\numberOfRepetitiveChanges{} scenarios, the edits are complex and context-dependent, meaning that a simple search/replace strategy is not enough to correctly apply the edits to all the necessary locations.

\paraheading{Experimental setup}
We use edits described in the \textit{diff} information of each scenario as examples of repetitive edits. To find the number of examples needed for \technique{} to perform all the repetitive edits in a commit, we start with a single edit, then run \technique{}. If only a subset of locations are found, we iteratively add more examples from the \textit{diff}. In this process, we prioritize variety, choosing examples that  cover different variations of the transformation.
If the edits performed by \technique{} and the edits in the diff do not match, we manually inspect them to check whether the developer missed a location or the locations were incorrectly edited.

\begin{table}[!t]
    \centering
    \footnotesize
    \caption{Summary of the evaluation on repetitive edits.
        Scope $=$ scope of the transformation; Ex $=$ number of examples; Dev.~$=$ number of locations modified by developers; \technique{}~$=$
        number of locations modified by \technique. Outcomes: \cmark\xspace~$=$ it performed the same edits as the developers; \starmark\xspace $=$ it performed
        more edits than the developers (manually validated as correct); \xmark\xspace $=$ it performed incorrect edits; ``---''\xspace $=$ it
    did not synthesize a transformation.}
    \begin{tabular}{ccccccc}
        \toprule
        Id & Project & Scope & Ex. & Dev. & \technique{} & Outcome \\
        \midrule
        1 & EF & Single file & 2 & 13 & 13 & \cmark \\
        2 & EF & Multiple files & 2 & 4 & 200 &\xmark\\
        3 & EF & Single file & 3 & 10 & 10 & \cmark \\
        4 & EF & Multiple files & 2 & 15 & 19 & \xmark \\
        5 & EF & Single file & 3 & 4 & 4 & \cmark \\
        6 & EF & Single file & 2 & 3 & 3 & \cmark \\
        7 & EF & Single file & 2 & 3 & 3 & \cmark \\
        8 & EF & Single file & 2 & 18 & 18 & \cmark \\
        9 & EF & Single file & 2 & 8 & 35 &\starmark\\
        10 & EF & Single file & 2 & 4 & 10 & \starmark \\
        11 & EF & Multiple files & 4 & 8 & 8 & \cmark \\
        12 & EF & Single file & 2 & 3 & 3 & \cmark \\
        13 & EF & Single file & 2 & 12 & 12 & \cmark \\
        14 & EF & Multiple files & 5 & 5 & 5 & \cmark \\
        15 & EF & Single file & 2 & 3 & 3 & \cmark \\
        16 & NuGet & Single file & 2 & 4 & 4 & \cmark \\
        17 & NuGet & Multiple files & 2 & 4 & 21 &\xmark\\
        18 & NuGet & Single file & 3 & 3 & 3 & \cmark \\
        19 & NuGet & Multiple files & 2 & 31 & 44 & \starmark\\
        20 & NuGet & Single file & 2 & 3 & 3 & \cmark \\
        21 & NuGet & Multiple files & 5 & 8 & 14 & \starmark\\
        22 & NuGet & Single file & 3 & 14 & 27 &\xmark\\
        23 & NuGet & Single file & 4 & 4 & 4 & \cmark \\
        24 & NuGet & Multiple files & 2 & 5 & 6 &\xmark\\
        25 & NuGet & Single file & 2 & 3 & 3 & \cmark \\
        26 & NuGet & Single file & 3 & 5 & 5 & \cmark \\
        27 & NuGet & Single file & 2 & 3 & 3 & \cmark \\
        28 & NuGet & Single file & 3 & 4 & 4 & \cmark \\
        29 & NuGet & Single file & 2 & 9 & 44 &\xmark\\
        30 & NuGet & Single file & 2 & 4 & 4 & \cmark \\
        31 & NuGet & Multiple files & 3 & 4 & 10 & \starmark \\
        32 & NuGet & Multiple files & 3 & 12 & 77 & \xmark \\
        33 & Roslyn & Single file & 3 & 3 & 21 & \starmark \\
        34 & Roslyn & Multiple files & 4 & 7 & 7 & \cmark \\
        35 & Roslyn & Multiple files & 3 & 17 & 18 & \starmark\\
        36 & Roslyn & Single file & 2 & 6 & 6 & \cmark \\
        37 & Roslyn & Single file & 2 & 9 & 9 & \cmark \\
        38 & Roslyn & Multiple files & 4 & 26 & 715 & \cmark \\
        39 & Roslyn & Single file & 3 & 4 & 4 & \cmark \\
        40 & Roslyn & Single file & 2 & 4 & 4 & \cmark \\
        41 & Roslyn & Single file & 6 & 14 & 14 & \cmark \\
        42 & Roslyn & Single file & 5 & 60 & --- & --- \\
        43 & Roslyn & Single file & 2 & 8 & 8 & \cmark \\
        44 & Roslyn & Multiple files & 3 & 15 & 15 & \cmark \\
        45 & Roslyn & Single file & 2 & 7 & 7 & \cmark \\
        46 & Roslyn & Single file & 4 & 13 & 14 & \cmark \\
        47 & Roslyn & Single file & 2 & 12 & 12 & \cmark \\
        48 & Roslyn & Single file & 2 & 4 & 4 & \cmark \\
        49 & Roslyn & Single file & 2 & 5 & 5 & \cmark \\
        50 & Roslyn & Single file & 2 & 3 & 3 & \cmark \\
        51 & Roslyn & Single file & 3 & 11 & 11 & \cmark \\
        52 & Roslyn & Single file & 2 & 5 & 5 & \cmark \\
        53 & Roslyn & Single file & 2 & 3 & 5 & \starmark \\
        54 & Roslyn & Single file & 3 & 5 & 5 & \cmark \\
        55 & Roslyn & Single file & 3 & 3 & 3 & \cmark \\
        56 & Roslyn & Single file & 3 & 6 & 6 & \cmark \\
        57 & Roslyn & Multiple files & 2 & 15 & 49 & \starmark\\
        58 & Roslyn & Single file & 2 & 4 & 7 & \starmark\\
        59 & Roslyn & Single file & 3 & 4 & 9 &\xmark \\
    \bottomrule
    \end{tabular}
    \label{tab:resultsStudy2}
\end{table}

\paraheading{Results}
Table~\ref{tab:resultsStudy2} summarizes our results.
In our evaluation, \technique{} synthesized transformations for \synthesizedCases{} out of \numberOfRepetitiveChanges{} scenarios.
In \numberOfRepetitiveChangesEqualsCommit{} (\percentOfRepetitiveChangesEqualsCommit\%) scenarios, the synthesized transformations applied the same edits applied by developers, whereas, in \numberOfRepetitiveChangesMoreCommit{} scenarios, the transformations applied more edits than developers did.
We manually inspected these scenarios, and conclude that \numberOfNeutralChanges{} transformations were correct (i.e., developers missed some edits).
We reported them to the developers of the respective projects.
So far, they confirmed \numberOfConfirmed{} of these scenarios.

In \numberOfIncorrect{} scenarios (17\%) the additional edits were incorrect, and revealed two limitations of the current DSL. First, some edits require further analysis to identify the location to apply them. For instance, in scenario 4, developers edited a local variable declaration. However, they did not perform the edit when this variable was reassigned to another object after its declaration. Extending our DSL to support this kind of operation would require some form of data flow analysis. The second limitation is related to our tree pattern matching. Some examples produced templates that were too general. For example, if two nodes have different numbers of children, we can currently only match them with respect to their type, which may be too general. To support
this kind of pattern, we plan to include additional predicates in our DSL such as \code{Contains}, which does not consider the entire list of children, but checks if any of the children matches a specific pattern.

Our technique, on average, required \examplesToEdit{} examples for synthesizing all transformations in a diff. The number of required examples may vary based on the examples selected by the developer. Additionally, changes in the ranking functions for giving preference to more general patterns over more restrict ones can also influence the number of examples. We leave a further investigation of example ordering and of our ranking system to future work.




\paraheading{Threats to validity}
With respect to construct validity, our initial baseline is the \textit{diff} information between commits. Some repetitive edits may have been performed across multiple commits, or developers may not have changed all possible code fragments. Therefore, there may be more similar edits in each scenario that were not considered. To reduce this threat, we manually inspect the additional edits applied by \technique{}. Concerning internal validity, the selection of example may influence the total number of examples needed to perform the transformation. As we mentioned, we prioritized examples that cover different patterns the transformation should consider. Finally, the sample of repetitive changes may not be representative for other kinds of software systems.

\section{Related Work}


\paraheading{Example-based program
transformations}
Meng et al.~\cite{ME13LASE,ME11SYST, rase} propose Lase, a technique for performing repetitive edits using examples.  Developers give two or more edited methods as examples, and Lase creates a context-aware abstract transformation. It uses clone detection and dependence analysis techniques to identify methods where the transformation should be applied and its context.
Lase only abstracts names of types, variables, and methods and can only find edits that mismatch with respect to these names. For instance, Lase cannot abstract the edits shown in Figures~\ref{fig:motivating1} and~\ref{fig:motivating2} since there are mismatches on expressions. Additionally, the edits in Lase have statement-level granularity, therefore limiting the type of expressible patterns. Finally, Lase cannot apply transformations that perform similar edits in the same method as shown in Figure~\ref{fig:motivating2}.
We plan to investigate the use of dependence-analysis to improve the quality of the transformations synthesized by \technique{}.


Other approaches allow expressing program transformations in a semi-automated way by using examples in combinations
with transformation templates~\cite{RO08EXAM,marti-chi07}.
Unlike these techniques our approach is fully automated.
Feser et al.~\cite{Feser:2015:SDS:2813885.2737977} propose a technique for synthesizing data-structure transformations from examples in functional programming languages.
Nguyen et al.~\cite{NG10AGRA} present LibSync, a technique that migrates APIs based on clients that already migrated. Tansey and Tilevich~\cite{TA08ANNO} present an example-based technique to migrate APIs that are based on annotations.
HelpMeOut~\cite{helpmeout} learn from examples transformations to fix compilation and run-time errors.
Unlike these techniques, \technique{} is not tailored to a specific domain.




Code completion techniques recommend code transformation to developers while
they are editing the source code.
Raychev et al.~\cite{RA13REFA} use data collected from large code repositories to learn likely code completions.
Similarly, Foster et al.~\cite{FO12WITC} use a large dataset of common code completions and recommend them to the user based on the code context.
Ge et al.~\cite{GE12RECO} propose a similar technique for auto-completing a refactoring manually started by the developer.
While these techniques are limited by the refactorings present in IDEs
and in the datasets, \technique{} can automate transformations that have never
been seen before.

\paraheading{Inductive programming}
Inductive programming (IP), also known as Programming-by-Example,  has been an active research area in the AI and HCI communities for over a decade~\cite{lieberman2001your}.  IP techniques have recently been developed for various domains including interactive synthesis of parsers~\cite{LeungSL15},
imperative data structure manipulations~\cite{storyboardfse}, and network policies~\cite{YuanAL14}. 
Recently, it has been successfully used in industry by FlashFill and FlashExtract~\cite{GU11AUTO,LE14FLAS,Mayer:2015:UIM:2807442.2807459}. FlashFill is a feature in Microsoft Excel 2013 that uses IP methods to automatically synthesize string transformation macros from input-output examples. FlashExtract is a tool for data extraction from semi-structured text files, deployed in Microsoft PowerShell for Windows 10 and as the Custom Field and Custom Log features in Operations Management Suite (a Microsoft log analytics tool).
The DSL of \technique{} is inspired by the ones of FlashExtract and FlashFill. While FlashFill uses the \code{ConstString} operator to create new strings and the \code{SubString} operator to get substrings from the input string, we use \code{NewNode} and \code{Reference} operators to compose the new subtree using new nodes or nodes from the existing AST. On the other hand, our DSL contains specific operators for performing tree edits and tree pattern matching.
FlashFill and FlashExtract gave rise to \prose{}, a novel framework of effective methods for IP~\cite{PO15FLAS}. While \prose{} has been
primarily used in the data wrangling domain, our technique shows its applicability to a novel unrelated domain -- learning program transformations.

\paraheading{Synthesis for education}
Singh et al.~\cite{autograder} propose AutoGrader, a program synthesis technique for fixing incorrect student submissions. Given a set of transformations that represent fixes for student mistakes (error-model) and an incorrect submission, AutoGrader uses symbolic execution to try all combinations of transformations to fix the student submission. While AutoGrader requires an error model, \technique{} automatically generates it from examples of fixes. In the future, we plan to use the symbolic search of AutoGrader to efficiently explore all transformations learned by \technique{}.

Rivers and Koedinger~\cite{Rivers2015} propose a data-driven technique for hint generation. The main idea is to generate concrete edits from the incorrect solution to the closest correct one. While they focus on comparing the entire AST, which can have many differences, our technique generalizes transformations that fix specific mistakes in student submissions. Kaleeswaran et al.~\cite{sumit-fse206} propose a semi-supervised technique for feedback generation. The technique clusters the solutions based on the strategies to solve it. Then instructors manually label in each cluster one correct submission. They formally validate the incorrect solutions against the correct one. Although our technique is completely automatic, we plan to investigate the use of formal verification to validate the transformations.

\paraheading{Program repair}
Automated program repair is the task of automatically changing
incorrect programs to make them meet a desired specification~\cite{GouForWei13}.
One of the main challenges is to efficiently search the space of all programs
to find one that behaves correctly.
The most prominent search techniques are enumerative or data-driven.
GenProg uses genetic programming
to repeatedly alter the incorrect program in the hope to make it correct~\cite{LeGDewForWei12}.
Data-driven approaches leverage large online code repositories
to synthesize likely changes to the input program~\cite{RayVecYah14}. Prophet~\cite{Long:2016:APG:2914770.2837617} is a patch generation system that learns a probabilistic application-independent model of correct code from a set of successful human patches.
Qlose provides ways to rank possible repairs based on a cost metric~\cite{qlose}.
Unlike these techniques, which use a global model of possible code transformations, \technique{} learns program-specific transformations using examples of code modification --- i.e., from both the original
and the modified program.

\section{Conclusions}
We presented \technique{}, a technique for
synthesizing syntactic  program transformations from examples.
Given a set of examples consisting of program edits,
\technique{} synthesizes a program transformation that is consistent with the examples.
Our synthesizer builds on the state-of-the-art program synthesis engine \prose{}.
To enable it, we develop (i) a novel DSL for representing program transformations, (ii) domain-specific constraints for the DSL operators,
which reduce the space of search for transformations, and (iii) ranking functions for transformation robustness, based on the structure of the synthesized transformations.
We evaluated \technique{} on two applications: synthesizing program transformations
that describe how students ``fix'' their programming assignments and
synthesizing program transformations that apply repetitive edits to large code bases.
Our technique learned program transformations that automatically fixed the program submissions of
 \averageHelpedStudents{}\% of the students participating in a large UC Berkeley class and it learned the transformations necessary to apply the correct code edits for
\refactoringAccuracy{}\% of the repetitive tasks we extracted from three large code repositories.

As future work, we plan to increase the expressiveness of our tree pattern expressions to avoid selecting incorrect locations due to over-generalization.
We aim at investigating the use of control-flow and data-flow analyses for identifying the context of the transformation, and the inclusion of negative examples and operators to specify undesired transformations.
In the context of HCI research, we want to design new user interaction models to enable instructors and developers to apply and debug the results of synthesized transformations.
In the education domain, we aim at developing new tools for providing personalized feedback for students based on the fixes learned by \technique{}.

In addition to being a useful tool, \technique makes two novel achievements in PBE.
First, it is the first application of backpropagation-based PBE methodology to a domain unrelated to data wrangling or string manipulation.
Second, in its domain it takes a step towards development of fully unsupervised PBE, as it automates extraction of input-output examples
from the datasets (that is, students' submissions or developers' modifications).
We hope that with our future work on incorporating flow analyses into witness functions, \technique will become the first major application
of inductive programming that leverages research developments from the entire field of software engineering.

  \newpage

\IEEEtriggeratref{45}


\bibliographystyle{IEEEtran}
\bibliography{ref}




\end{document}